\documentclass[aps, prl, twocolumn, showpacs, superscriptaddress]{revtex4-2}  

\usepackage{epsfig}
\usepackage{graphicx}
\usepackage{amsmath}
\usepackage{amsfonts}
\usepackage{amssymb}
\usepackage{xcolor}

\begin{document}

\title{Can the ``shadow'' of graphene band clarify its flatness?}

\author{Matteo Jugovac}
\affiliation {Istituto di Struttura della Materia, Consiglio Nazionale delle Ricerche, Area Science Park, I-34149 Trieste, Italy}
\affiliation {Elettra Sincrotrone Trieste, Strada Statale 14 km 163.5, 34149 Trieste, Italy}

\author{Cesare Tresca}
\affiliation {Dipartimento di Scienze Fisiche e Chimiche, Universit\`{a}  dell'Aquila, Via Vetoio 10, 67100, L'Aquila, Italy}
\affiliation {CNR-SPIN L'Aquila, Via Vetoio 10, 67100 L'Aquila, Italy}

\author {Iulia Cojocariu}
\affiliation {Peter Gr\"{u}nberg Institute (PGI-6), Forschungszentrum J\"{u}lich GmbH, 52425 J\"{u}lich, Germany}

\author {Giovanni Di Santo}
\affiliation {Elettra Sincrotrone Trieste, Strada Statale 14 km 163.5, 34149 Trieste, Italy}

\author {Wenjuan Zhao}
\affiliation {Elettra Sincrotrone Trieste, Strada Statale 14 km 163.5, 34149 Trieste, Italy}

\author {Luca Petaccia}
\affiliation {Elettra Sincrotrone Trieste, Strada Statale 14 km 163.5, 34149 Trieste, Italy}

\author{Paolo Moras}
\affiliation {Istituto di Struttura della Materia, Consiglio Nazionale delle Ricerche, Area Science Park, I-34149 Trieste, Italy}

\author {Gianni Profeta}
\affiliation {Dipartimento di Scienze Fisiche e Chimiche, Universit\`{a}  dell'Aquila, Via Vetoio 10, 67100, L'Aquila, Italy}
\affiliation {CNR-SPIN L'Aquila, Via Vetoio 10, 67100 L'Aquila, Italy}

\author {Federico Bisti}
\email{federico.bisti@univaq.it}
\affiliation {Dipartimento di Scienze Fisiche e Chimiche, Universit\`{a}  dell'Aquila, Via Vetoio 10, 67100, L'Aquila, Italy}

\normalsize
\date{\today}

\begin{abstract}
Graphene band renormalization at the proximity of the van Hove singularity (VHS) has been investigated by angle-resolved photoemission spectroscopy (ARPES) on the Li-doped quasi-freestanding graphene on the cobalt (0001) surface. The absence of graphene band hybridization with the substrate, the doping contribution well represented by a rigid energy shift and the excellent electron-electron interaction screening ensured by the metallic substrate offer a privileged point of view for such investigation. A clear ARPES signal is detected along the M point of the graphene Brillouin zone, giving rise to an apparent flattened band. By simulating the graphene spectral function from the density functional theory calculated bands, we demonstrate that the photoemission signal along the M point originates from the ``shadow'' of the spectral function of the unoccupied band above the Fermi level. 
Such interpretation put forward the absence of any additional strong correlation effects at the VHS proximity, reconciling the mean field description of the graphene band structure even in the highly doped scenario.
\end{abstract}


\maketitle

Superconducting phase in the twisted graphene bilayer\cite{Cao2018} has strongly renewed the interest on flat band materials, in which a nearly undispersed (flat) energy band is present in a relevant portion of the Brillouin zone.
A material presenting a flat band in the proximity of the Fermi level (E$_F$) is more inclined to manifest exotic electronic phases since any whatever small electronic interactions could be strongly enhanced by the divergent density of the states (DOS) coming from this low dispersing band.
Therefore, the presence of such DOS singularity is the source for strong electronic instabilities able to open a gap near E$_F$ possibly developing a new symmetry breaking ground state. Those instabilities can drive the system to develop magnetic orders\cite{Mielke1991, Tasaki1992, Mielke1993, Honerkamp2008, Pamuk2017, Calandra2018, Tresca2019, Campetella2020}, superconducting states\cite{Nandkishore2012,Honerkamp2008,DaSilvaNeto2014,Comin2014,Cao2018,Lu2019} or charge density wave phases\cite{Honerkamp2008}.

The strategies to get DOS singularity are many, and graphene holds enormous potential in this field.  
It emerges from graphene Moir\'e  physics at magic angles in connection with a flat band originated by Dirac bands hybridization in bilayer system, as in twisted graphene bilayers\cite{Cao2018, Kerelsky2019, Lu2019, Lisi2021} or in epitaxial multilayer graphene\cite{Pierucci2015, Henck2018, Henni2016, Pamuk2017, Marchenko2018,Baima2018, Campetella2020}.  
In addition, a DOS singularity is already present in the band structure of graphene at a saddle point in the unoccupied energy region far from the Dirac point (neutrality), known as van Hove singularity (VHS)\cite{CastroNeto2009}. Thus, an efficient alternative route to reach such DOS singularity is to over-dope the quasi-freestanding monolayer of graphene bringing E$_F$ at VHS\cite{McChesney2010,Bisti2015,Khademi2016,Verbitskiy2016,Tresca2018,Link2019,Antoniazzi2020}.
However, since it was considered too far in energy from the Dirac point to be accessible by chemical doping or gating, much of the attention shifted to twisted systems\cite{Li2010}.
Only in recent years, improving in graphene growth on appropriate substrates, deposition techniques\cite{Rosenzweig2020} and band structure measurement, allowed to ''slowly'' increase the amount of charge deposited on monolayer graphene, demolishing the belief that the VHS in graphene could not be reached with present technology.

Recent band structure measurement of doped graphene have detected relevant deviations from a simple rigid shift of neutral graphene or of the first-principles density functional theory (DFT) band structure. Apart from the general consensus on the band renormalization around 170 meV due to the electron-phonon interaction\cite{Bisti2021}, a strong flattening of the band close to E$_F$ has been interpreted as originating from electron-electron correlations due to on-site Coulomb interaction U\cite{McChesney2010} or spin fluctuations\cite{Link2019}. However, the interpretation of the experimental results are complicated by the graphene's band hybridization with dopant atoms and/or intercalant atoms deposited on the substrate.
A recent example is represented by the Yb 4f-orbital anticrossing-type hybridization with the graphene $\pi^*$-band\cite{Rosenzweig2020}, or by the flat band formation by hybridization in heavily Cs-doped graphene\cite{Ehlen2020}.
Unfortunately, the desirable conditions to observe the VHS in (pure) graphene seem mutually exclusive: growth a nearly free-standing graphene, but with negligible interaction with the substrate; use of metallic substrates to screen the electron-electron interaction through  the substrate; heavily dope graphene to reach the VHS point, but avoiding  hybridization with dopant electronic states to preserve graphene band structure.

In this manuscript, we successfully realized most of the requirements, reporting data on the electronic structure of heavily Li-doped graphene on Co(0001) probed by ARPES and predicted by first-principles calculations. We observed graphene bands over an unprecedented large energy range, excellently described by density functional theory with local energy functionals and marginal renormalization of the bandwidth. We demonstrate that the presence of spectral signal along the $\Gamma$-M direction, resembling an apparent strongly renormalized (flattened) band close to E$_F$, can be instead clearly reproduced by DFT bands with a careful simulation of the spectral function. 
Thus the only detectable many-body interaction still holding in graphene is represented by the electron-phonon coupling, which we observed as a kink in the band dispersion.

\begin{figure}[t]
 \includegraphics [width=8cm] {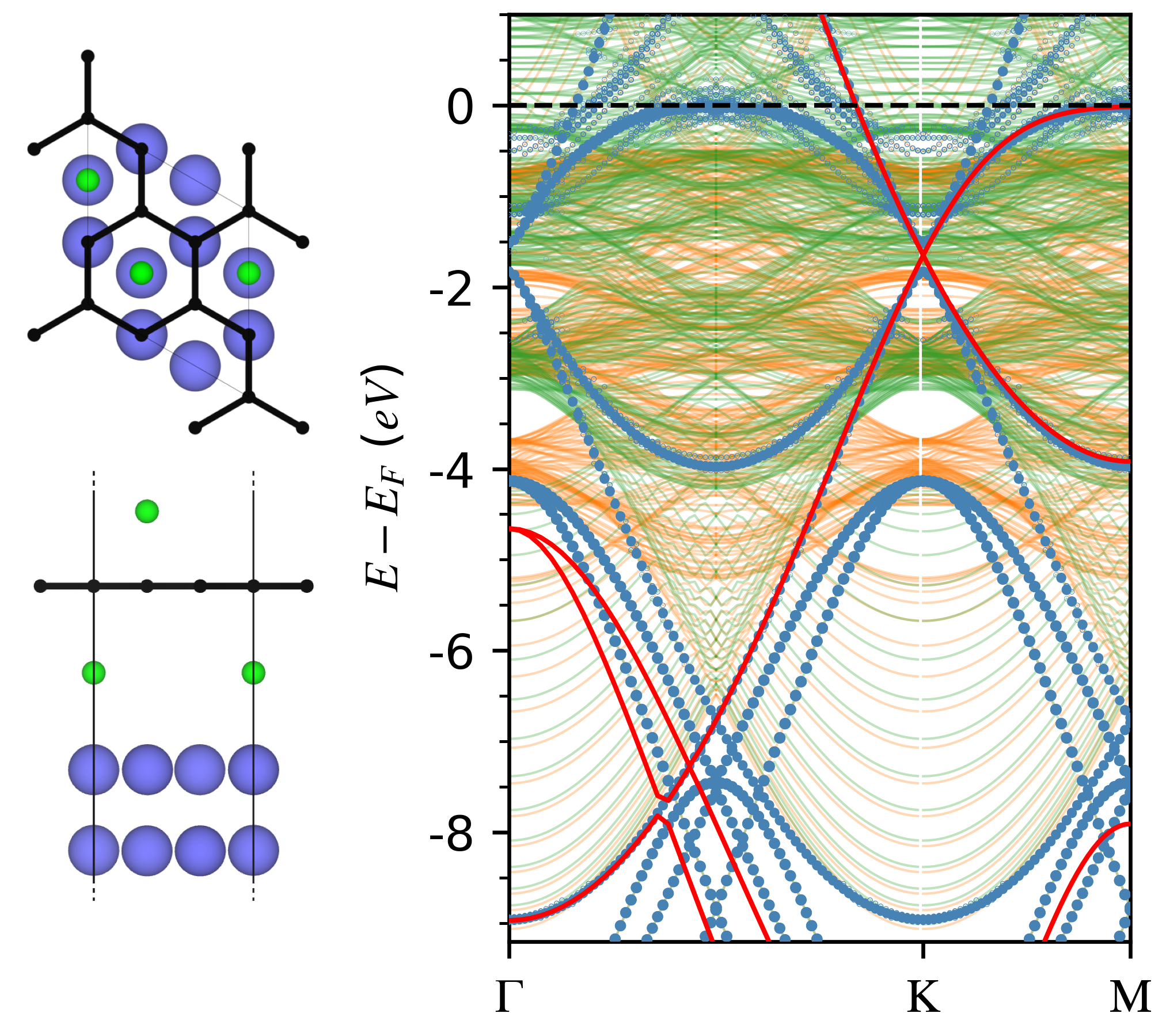}
 \caption {On the left, atomic model of Li (green) intercalated graphene (black) supported by the Co substrate (blue). On the right, the spin-up (green) and spin-down (orange) electronic band structure for the overall system. Blue dots represents the carbon $pz$-orbital weight for the states. Red line is the electronic band structure for an isolated graphene layer.}
 \label {fig_dft}
\end{figure}

Pristine monolayer (epitaxial (1x1)) graphene samples were prepared \textit{in-situ} under ultra-high vacuum conditions by chemical vapor deposition (CVD) on Co(0001) thin films (about 10 nm) epitaxially grown on W(110), using ethylene as carbon precursor\cite{Jugovac2019}. To decouple graphene from the Co substrate (making it quasi-freestanding) lithium was evaporated (while keeping the sample at 80 K) from commercial SAES metal dispensers and intercalated at room temperature (see Supplementary Material for the graphene growth Li intercalation process \cite{SuppMat}). From this doping level, we successfully achieved additional doping further depositing lithium while keeping the sample temperature at 80 K, thus sandwiching graphene between two Lithium monolayers (see below).
ARPES data were collected at the BaDElPh\cite{Petaccia2009} (VUV-Photoemission) beamline of the Elettra Synchrotron (Trieste, Italy) using 33 (40) eV photon energy, keeping the sample at 80 (20) K. The total energy and angular resolutions were set to 20 meV (full-width-at-half-maximum of the Gaussian model fitting the experimental Fermi edge) and 0.1°, respectively. 

We predicted the electronic structure of this heavily doped graphene, modeling it by first-principles DFT (see Supplementary Material for computational details \cite{SuppMat}).

The alkali metal atoms were adsorbed (and intercalated) on the hollow site of graphene\cite{Profeta2012} in a 
$\sqrt{3}\times\sqrt{3}$-R30$^\circ$ 
reconstruction (see Fig.\ref{fig_dft} for a representation of the  structural model).
The predicted band structure is reported in Fig. \ref{fig_dft} (see also Fig. S3 of the Supplementary Material for the system without additional Li on top of graphene\cite{SuppMat}). The carbon projected states (blue points) are nearly indistinguishable from free-standing graphene's band structure (red line), indicating that this structural model realises a  completely decoupled graphene from the substrate. In addition, the magnetic exchange field of the substrate (which is spin-polarized) is not affecting the graphene. 
But, more important, the reached doping level is such that the van Hove singularity at the M point should be occupied.

\begin{figure}[t]
 \includegraphics [width=8cm] {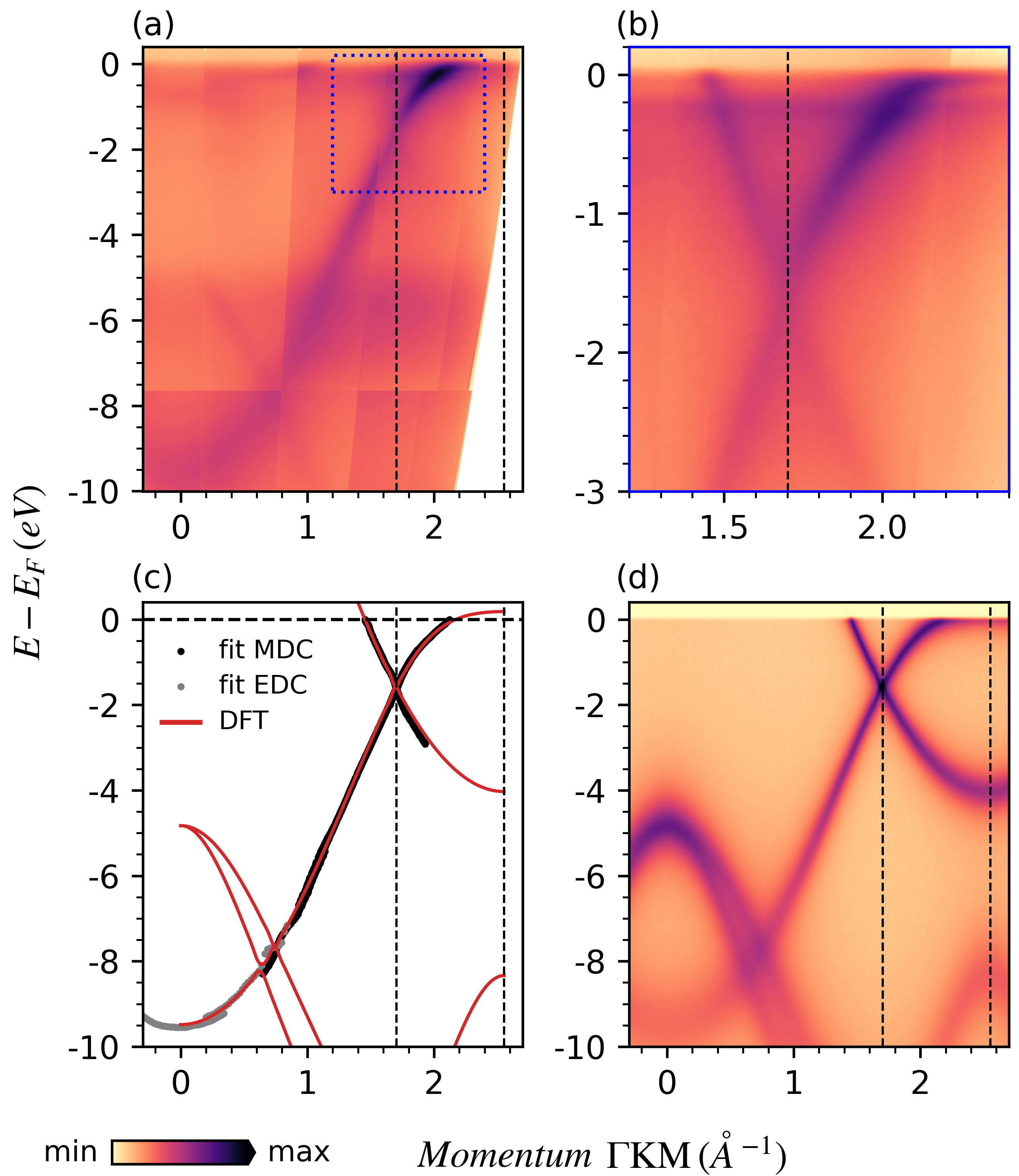}
 \caption {Graphene Dirac cone dispersion along $\Gamma$KM direction of the Li/Gr/Li/Co system. (a), ARPES spectra using p-polarization. (b), linear combination of the s- and p-polarization ARPES signals in the proximity of the Dirac point. (c), extracted band dispersion from MDC and EDC analysis (black and grey dots respectively) are reported with the stretched (by 1.08 factor) PBE-sol DFT calculations. (d), simulated spectral function derived from stretched DFT theoretical dispersion.}
 \label {fig_cone}
\end{figure}

At this point, the experimental confirmation is crucial.
In Fig. \ref{fig_cone}, we report the ARPES overview of the Dirac cone for the sample after a first step of  Li-intercalation (leading only to the graphene/substrate decoupling) and a further deposition step to reach the highest possible doping level. The Dirac point results to be shifted down to -1.58 eV. It is important to note that we are not able to distinguish any discontinuity on both $\pi$-band branches at the Dirac point, thus we exclude any relevant hybridization of the graphene with the Li atoms or with the substrate, meeting the first two requirements listed in the introduction. Lithium is indeed able to detach the graphene layer, as suggested by our theoretical calculation, and in analogy to what obtained on the same substrate with Si \cite{Usachov2018} and O \cite{Jugovac2020} intercalation. However, lithium brings further doping. 
The observed overall doping is quite high but the actual occupancy of the van Hove singularity is not as obvious. Apart from the dominant graphene spectral signal, a small and flat feature is detected at around 0.2 eV, which we attribute to emission from the 3d bands of the cobalt substrate (\emph{See Supplementary Material}). However, we cannot exclude the possibility of the presence of a flat band feature observed in other similar systems at the VHS proximity at similar binding energies, which was ascribed to polaron formation due to the coupling with optical phonons \cite{McChesney2010, Link2019, Rosenzweig2020} (see Supplementary Material on the possible presence of the polaron band). 
In order to gain a more quantitative analysis of the graphene $\pi$-band dispersion, the momentum (energy) dispersion curves, MDC (EDC), have been analyzed for the high (low) dispersing part of the $\pi$-bands. The data are reported in Fig. \ref{fig_cone}(c) along with the results of PBEsol-DFT calculation on isolated graphene. In order to match the experimental data, a marginal renormalization of the theoretical bandwidth is needed by a factor of 1.08, thank to the highly metallic character of the substrate which guarantees an excellent screening of the Coulomb interaction (which strongly renormalizes graphene bands when grown on insulating substrates\cite{Hwang2012}).
Within this picture, the unoccupied part of the $\pi^*$-band around the M point is extremely close to the Fermi level (within 0.18 eV), resulting in a "flat signal" extending through the M point (visible as a shadow at the E$_F$), which can thus find a natural origin in the energy broadening stemming from the photoemission final state lifetime. Other additional many-body effects, beyond the mean-field DFT, are not required to explain the graphene band structure. 
The correlation between the finite electronic final state lifetime value with the unavoidable energy broadening of the probed signal is an explicit manifestation of the particle-wave duality nature of electrons, as described by the Bohr derivation of the uncertainty principle \cite{Bohr1928}. 
To demonstrate this effect, in Fig. \ref{fig_cone}(d) we report the simulated photoemission signal as Lorentzians centered at the DFT eigenvalues ($Eb$), with an energy-dependent width as $\gamma = \gamma_0 + \gamma_1 \cdot Eb \:$ with $\gamma_0 = 0.15 \: eV$ and $\gamma_1 = 0.01$, tuned from the experimental EDC data analysis. The agreement with the experimental curve is indeed very good.

\begin{figure}[t]
 \includegraphics [width=8.5cm] {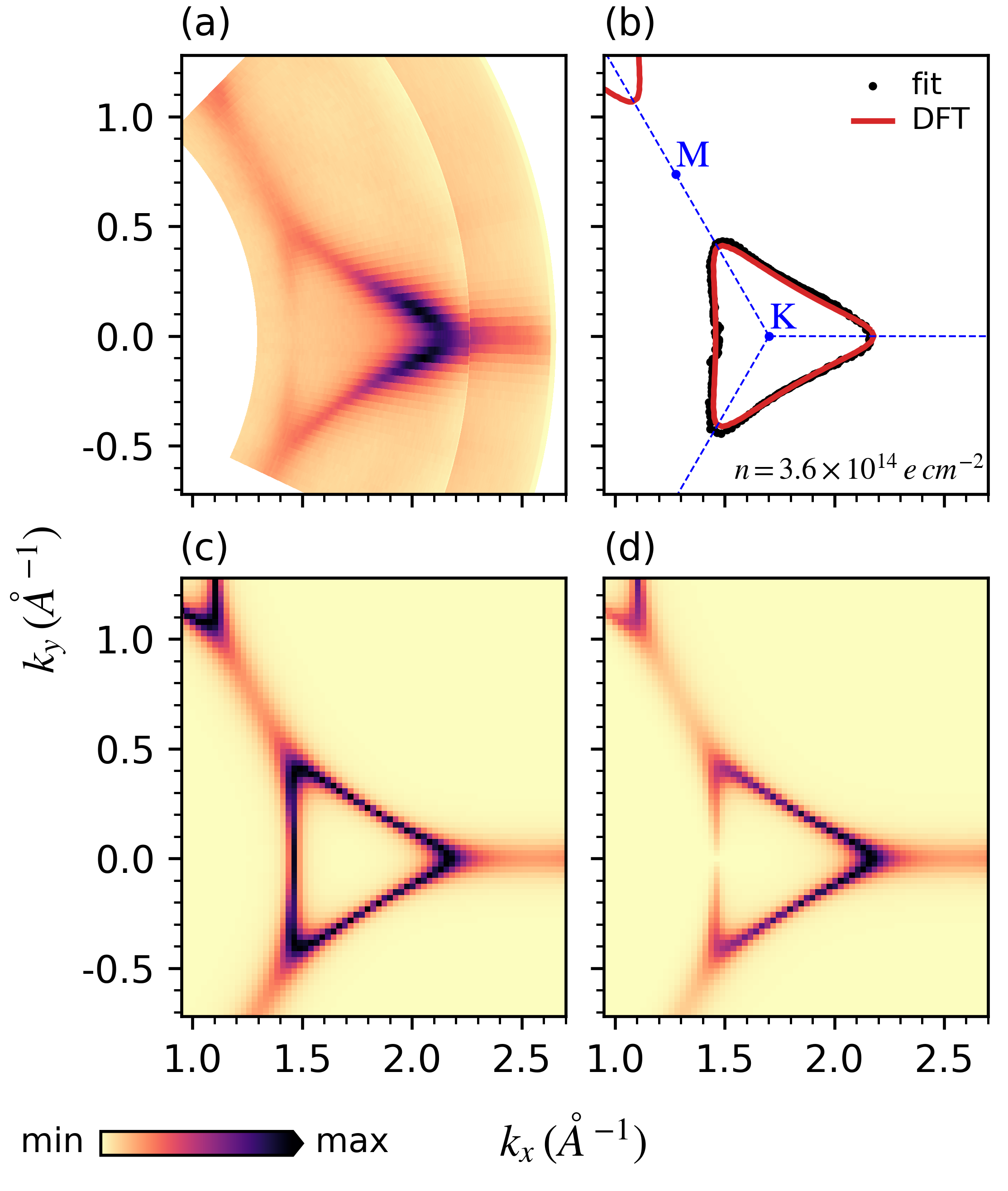}
 \caption {Fermi surface of the Li/Gr/Li/Co system. (a), ARPES spectra using p-polarization. (b), extracted band dispersion from MDC analysis (black dots) is reported along with the stretched (by a 1.08 factor) PBE-sol DFT calculations (red curve). Simulated spectral function derived from the stretched DFT theoretical dispersion in (b) without (c) and with (d) the light polarization selection rules.}
 \label {fig_fs}
\end{figure}

A further convincing proof comes from the simulation of the Fermi surface as reported in Fig. \ref{fig_fs}. The MDC analysis of the ARPES signal, reported in Fig. \ref{fig_fs}(a), is exactly matched by the (renormalized) DFT modes as shown in \ref{fig_fs}(b) (the excellent agreement with the theoretical model is also demonstrated in Fig. S5 of the Supplementary Material for different iso-energy maps \cite{SuppMat}). The shadow extending along the M point, reported in \ref{fig_fs}(c), is perfectly reproduced by the simulated spectra. 
To increase even more the simulation fidelity, in Fig. \ref{fig_fs}(d) we considered also the effect of light polarization in the selection rules.
In the isolated graphene case, eigenfunctions are symmetric or antisymmetric with respect to the mirror plane passing between the two carbon atoms of the graphene unit cell. By considering the final state as symmetric with respect to this mirror plane (as in our case, or in general for pure free-electron final state), then depending on the  light polarization symmetry with respect to this plane, we select photoemission from symmetric or antisymmetric initial states\cite{Bisti2017}. 
The simulated spectra is then derived considering this weight in the signal intensity (see Supplementary Material for its exact evaluation\cite{SuppMat}). The net result is a vanishing signal in the middle of one side of the Fermi surface triangle, clearly observed in the experiment (Fig. \ref{fig_fs}(a)) and perfectly predicted by the simulation in Fig. \ref{fig_fs}(d) (same effect is obviously acceptable for the analogue iso-energy map below the Dirac point at -3.7 eV, as shown in Fig. S6 of the Supplementary Material\cite{SuppMat}). The observed polarization effects is a solid confirmation of the absence of any relevant additional interaction on top of graphene since its fundamental mirror plane symmetry is not destroyed (by doping) but perfectly intact, again pointing to the realization of free-standing heavily doped graphene.

From the Fermi surface area extension we can estimate the effective electron doping of about $3.6 \times 10^{14} e/cm^{2}$, a value comparable with what reported in the recent literature\cite{McChesney2010, Link2019}. In particular,  McChesney \textit{et al.} \cite{McChesney2010} decorated graphene with calcium above and below the graphene sheet and by depositing additional potassium atoms, were able to shift even more down the Dirac point, at least by 0.1 eV. From what is possible to extract from their Figure 1, the new Fermi energy enhances even more the shadow effect proposed in our interpretation; but still without reaching (exactly) the VHS.

These findings demonstrate that strong correlation effects due to electron-electron interaction, invoked to justify the presence of extended van Hove singularities on the Fermi surface of alkali metals- \cite{McChesney2010}, Gd-doped graphene\cite{Link2019} and Yb-doped \cite{Rosenzweig2020}, could be not necessary to understand the experimental band structure of heavily electron-doped graphene.
In addition, such weak interaction (compared to the fermionic band width) scenario does not preclude the presence of possible non trivial phases in our system. Indeed, the theorized non trivial phases like $d$-wave superconductivity \cite{Nandkishore2012} is based on the assumption of a weak interaction in graphene. Therefore our system, being close to DOS singularity, still holds the promise for hosting such topological superconductivity phase.

\begin{figure}[t]
 \includegraphics [width=8.4cm] {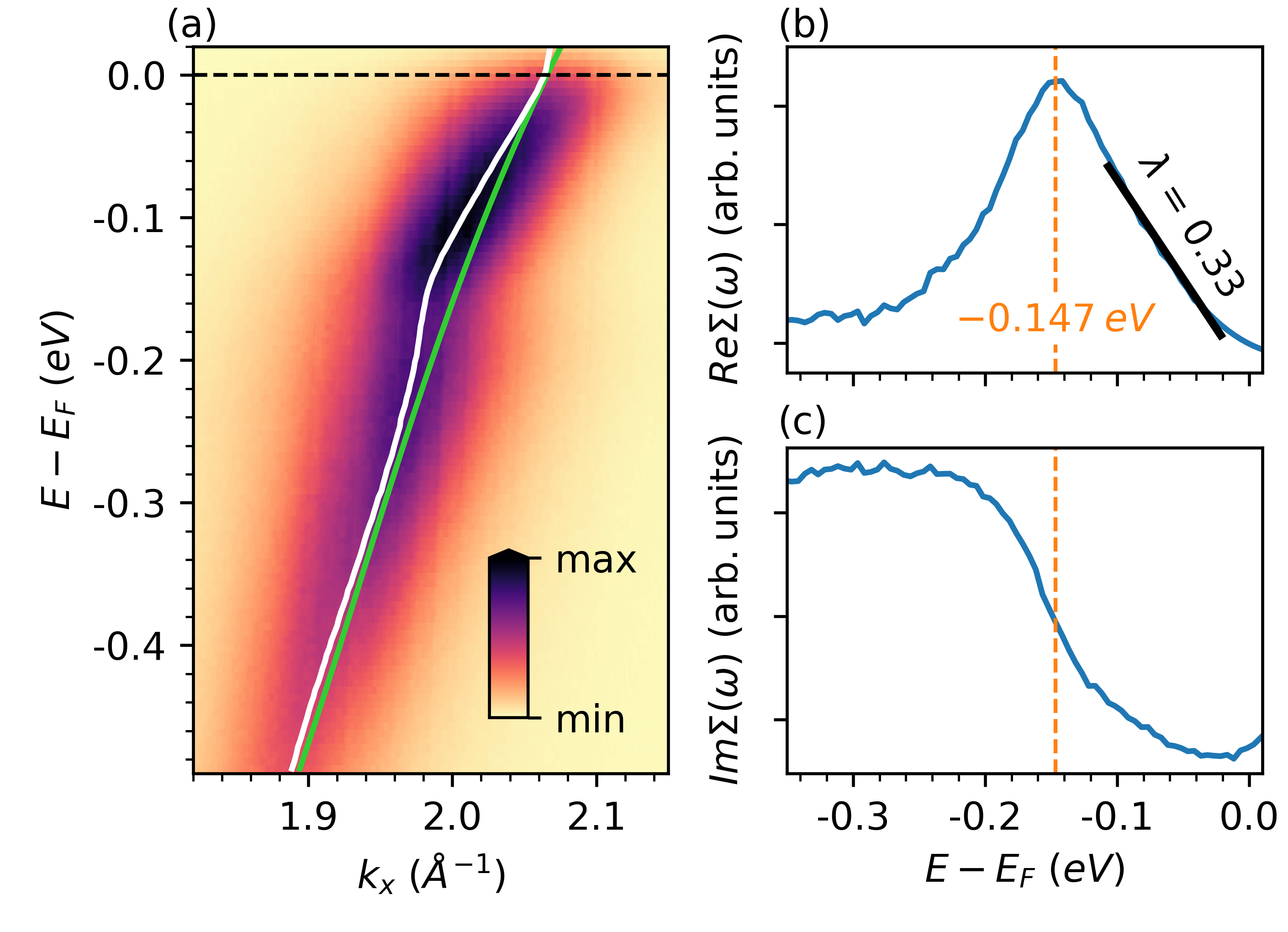}
 \caption {(a), high resolution ARPES spectra of graphene band dispersion along the $\Gamma KM$ direction with the extracted data from MDC analysis (white lines) and the bare bands resulting from the self-consistent analysis (green lines). Self-energy real (b) and imaginary (c) part.}
 \label {fig_kink}
\end{figure}

Looking at the experimental ARPES spectra, the more evident deviation form the mean-field band structure is the well known kink feature at around -170 meV, which represents the only manifestation of many-body effects  relevant near the Fermi level even at the proximity of the van Hove singularity.
In Fig. \ref{fig_kink} we report the kink analysis for the present system, using the same computational analysis as in Ref. \cite{Bisti2021}. In this case, in order to reduce the ``shadow'' signal near the Fermi energy (which can interfere with the fitting procedure), we used a  slightly under-doped ($2.5 \times 10^{14}/cm^{2}$) system with respect the data shown before (see Fig. S2(g) of the Supplementary Material and related discussion \cite{SuppMat}), considering that, for the purpose of this analysis the overall doping is not so relevant. 
Interestingly, we notice an unexpected significant energy shift of the kink with respect to the value of highly doped system with Li (-169 meV)\cite{Bisti2021},
which is now centered at -147 meV, representing the characteristic E$_{2g}$ phonon frequency. 
This shift  is even larger than that observed in a graphene completely substituted with ${}^{13}$C (-162 meV)\cite{Bisti2021} and cannot be explained by an artificial error due to data manipulation. 
The observed softening of the phonon modes is a natural consequence of the effect of Li decoration on both sides of graphene and of the increased doping level: indeed first- principles theoretical calculations of phonon dispersion for a graphene layer decorated on both sides, have correctly predicted this softening (see Supplementary Material of Ref. \cite{Profeta2012}). 
This represent a further, indirect, confirmation of the physical realization of a heavily doped ideal free-standing graphene. 

In conclusion, Li decoupled and highly doped graphene on Co(0001), resulted to be an excellent realization of an ideal free-standing graphene layer in highly doped regime, without spurious effects as substrate or dopant interaction. From the analysis of its band structure we were able to demonstrate, at the same time that $i)$ a single electron picture is capable of explaining the ARPES signal only including a relatively small energy renormalization, but without invoking strong correlation effects to induce band structure flattening in the proximity of the van Hove singularity, $ii)$ ARPES spectra can show signal from the unoccupied band if sufficiently close to the Fermi level within the thermal broadening of the Fermi distribution and the life-time of the photoemission hole, $iii)$ the only detectable many-body feature near the Fermi level is the electron-phonon kink and $iv)$ the present system shows a phonon softening induced by the alkali metals decoration on both sides of graphene. 

The authors acknowledge Elettra Sincrotrone Trieste for providing access to its synchrotron radiation facilities.
G. P. wishes to acknowledge financial support from the Italian Ministry for Research and Education through PRIN-2017 project ``Tuning and understanding Quantum phases in 2D materials - Quantum 2D" (IT-MIUR Grant No. 2017Z8TS5B).

\bibliography{graphene}

\end{document}